\documentclass[11pt]{article}
\usepackage{amsmath}
\usepackage{amssymb}
\usepackage{amsthm}
\usepackage{xcolor}
\usepackage{caption}
\usepackage{calc}
\usepackage{hyperref}
\usepackage{graphicx}
\usepackage{wrapfig}
\usepackage{enumitem}
\usepackage{cite}
\usepackage{geometry}
\usepackage{amsmath}
\usepackage{graphicx}
\usepackage{algorithm}
\usepackage[noend]{algpseudocode}
\usepackage{natbib}
\usepackage{url} 

\newcounter{algsubstate}
\renewcommand{\thealgsubstate}{\alph{algsubstate}}
\newenvironment{algsubstates}
  {\setcounter{algsubstate}{0}%
   \renewcommand{\State}{%
     \stepcounter{algsubstate}%
     \Statex {\footnotesize\thealgsubstate:}\space}}
  {}

\begin{document}


\title{Eigenvectors from Eigenvalues Sparse Principal Component Analysis (EESPCA)}

\author{H. Robert Frost$^{1}$}
\date{}
\maketitle
\begin{center}
\textit{
$^1$Department of Biomedical Data Science\\
Geisel School of Medicine \\
Dartmouth College \\
Hanover, NH 03755, USA \\
rob.frost@dartmouth.edu
}
\end{center}

\begin{abstract}
We present a novel technique for sparse principal component analysis. This method, named Eigenvectors from Eigenvalues Sparse Principal Component Analysis (EESPCA), is based on the formula for computing squared eigenvector loadings of a Hermitian matrix from the eigenvalues of the full matrix and associated sub-matrices. We explore two versions of the EESPCA method: a version that uses a fixed threshold for inducing sparsity and a version that selects the threshold via cross-validation. Relative to the state-of-the-art sparse PCA methods of Witten et al., Yuan \& Zhang and Tan et al., the fixed threshold EESPCA technique offers an order-of-magnitude improvement in computational speed, does not require estimation of tuning parameters via cross-validation, and can more accurately identify true zero principal component loadings across a range of data matrix sizes and covariance structures. Importantly, the EESPCA method achieves these benefits while maintaining out-of-sample reconstruction error and PC estimation error close to the lowest error generated by all evaluated approaches. EESPCA is a practical and effective technique for sparse PCA with particular relevance to computationally demanding statistical problems such as the analysis of high-dimensional data sets or application of statistical techniques like resampling that involve the repeated calculation of sparse PCs.
\end{abstract}

\noindent%
{\it Keywords:}  principal component analysis, sparse principal component analysis, sparse eigenvalue decomposition, eigenvector-eigenvalue identity
\vfill

\section{Introduction}
\label{sec:intro}

\subsection{Principal component analysis (PCA)} \label{sec:PCA}
PCA is a widely used statistical technique that was developed independently by Karl Pearson \citep{Pearson:1901fk} and Harold Hotelling \citep{Hotelling:1933uq} in the early part of the 20th century. PCA performs a linear transformation of multivariate data into a new set of variables, the principal components (PCs), that are linear combinations of the original variables, are uncorrelated and have sequentially maximum variance \citep{jolliffe2002principal}. PCA can be equivalently defined by the set of uncorrelated vectors that provide the best low-rank matrix approximation, in a least squares sense.
The solution to PCA is given by the eigenvalue decomposition of the sample covariance matrix with the variance of the PCs specified by the eigenvalues and the PC directions defined by the eigenvectors. 

Because PCA is defined in terms of the eigenvectors and eigenvalues of the sample covariance matrix, it is related to a wide range of matrix analysis methods and multivariate statistical techniques with an extremely large number of applications \citep{jolliffe2002principal, Jolliffe:2016aa}. Although most commonly used for unsupervised linear dimensionality reduction and visualization, PCA has been successful applied to statistical problems including regression (e.g., principal components regression\citep{Hastie:2009qf}), clustering (e.g., use of PCs as input for clustering methods \citep{Waltman_2013, Hafemeister:2019fr}), and non-linear dimensionality reduction (e.g., seed directions for non-linear methods \citep{lel2018umap}).
In the biomedical domain, PCA has been extensively employed for the analysis of genomic data including measures of DNA variation, DNA methylation, RNA expression and protein abundance \citep{Ma:2011kl}. Common features of these datasets, and the motivation for eigenvalue decomposition methods, are the high dimensionality of the feature space (i.e., from thousands to over one million), comparatively low sample size (i.e., $p \gg n$) and significant collinearity between the features. The most common uses of PCA with genomic data involve dimensionality reduction for visualization or clustering \citep{Stuart:2019yq}, 
with population genetics an important use case \citep{Patterson:2006uq}. PCA has also been used as the basis for feature selection \citep{Lu:2011ys}, and gene clustering \citep{kluger_spectral_2003}. More recent applications include gene set enrichment of bulk or single cell gene expression data \citep{Tomfohr:2005bv, Fan:2016aa}.

\subsection{Mathematical notation for PCA} \label{sec:PCA_notation}

Let $\mathbf{X}$ be an $n \times p$ matrix that holds $n$ independent samples drawn from a $p$-dimensional joint distribution with population covariance matrix $\boldsymbol{\Sigma}$. Without loss of generality, we can assume the columns of $\mathbf{X}$ are mean-centered. The unbiased sample covariance matrix is therefore given by $\hat{\boldsymbol{\Sigma}} = 1/(n-1) \mathbf{X}^T \mathbf{X}$. PCA can be performed via the direct eigenvalue decomposition of $\hat{\boldsymbol{\Sigma}}$ with the eigenvalues $\lambda_i$ equal to the PC variances and the unit length eigenvectors $\mathbf{v_i}$ equal to the PC loadings. For computational reasons, PCA is more commonly performed via the singular value decomposition (SVD) of $\mathbf{X}$:
$\mathbf{X} = \mathbf{U} \mathbf{D} \mathbf{V}^T$,
where $\mathbf{U}$ and $\mathbf{V}$ are both orthonormal matrices (i.e., $\mathbf{U}^T\mathbf{U} = \mathbf{V}^T\mathbf{V}=\mathbf{I}$), the columns of $\mathbf{V}$ represent the PC loading vectors, the entries $d_i$ in the diagonal matrix $\mathbf{D}$, arranged in decreasing order, are proportional to the square roots of the PC variances ($d_i = (n-1)\sqrt{\lambda_i}$) and the columns of $\mathbf{U} \mathbf{D}$ are the principal components. Consistent with the optimal low-rank matrix approximation property of PCA, the first $r$ components of the SVD minimize the squared Frobenius norm between the original $\mathbf{X}$ and a rank $r$ reconstructed version of $\mathbf{X}$. Specifically, if $\mathbf{U}_r$, $\mathbf{D}_r$ and $\mathbf{V}_r$ hold the first $r$ columns of $\mathbf{U}$, $\mathbf{D}$ and $\mathbf{V}$, $\hat{\mathbf{X}}_r = \mathbf{U}_r \mathbf{D}_r \mathbf{V}_r^T$  is the SVD rank $r$ reconstruction of $\mathbf{X}$, which minimizes $||\mathbf{X} - \hat{\mathbf{X}}_r||^2_F$ for all possible rank $r$ reconstructions.

In the remainder of the manuscript we will use $||\mathbf{x}||_0$ to refer to the $l_0$ norm of vector $\mathbf{x}$ (i.e., number of non-zero values), 
$||\mathbf{x}||_2$ to refer to the Euclidean or $l_2$ norm, 
and $\mathbf{x} \odot \mathbf{y}$ to refer to the element-wise multiplication of vectors $\mathbf{x}$ and $\mathbf{y}$. In general, we will use bold font for matrices and vectors and non-bold for scalars. Additional mathematical notation is defined when first used.

\subsection{Running example} \label{sec:example}

To help illustrate the concepts discussed in this paper, we introduce a simple example data set based on a 10-dimensional multivariate normal (MVN) distribution. 
The R logic needed to reproduce all of the results for this example can be found in the Supplementary Material.
The population mean is set to the zero vector, $\boldsymbol{\mu} = (0,0,0,0,0,0,0,0,0,0)$,
and the population covariance matrix is given a two block covariance structure with a covariance of 0.5 among the first four variables and between the last two variables, and a covariance of 0 among all other variables.
All population variances are set to one, which aligns with the common practice of performing PCA after standardization.
For this $\boldsymbol{\Sigma}$, the first population PC has equal non-zero loadings for just the first four variables and the second population PC has equal non-zero loadings for just the last two variables. If the matrix $\mathbf{V}$ holds the population PC loadings, the first two columns are:
\begin{align*} 
	 \mathbf{V[,(1,2)]^T} = \begin{bmatrix}
	   -0.5 & -0.5 & -0.5 & -0.5 & 0 & 0 & 0 & 0 & 0 & 0 \\
	   0 & 0 & 0 & 0 & 0 & 0 & 0 & 0 & 0.7071068 & 0.7071068 
	 \end{bmatrix}
\end{align*}
\noindent The variances of these population PCs are $\boldsymbol{\lambda} = (2.5, 1.5)$.

For an example set of 100 independent samples drawn from this MVN distribution, the loadings of the first two sample PCs (rounded to three decimal places) are:
\begin{align*} 
	 \mathbf{\hat{V}[,(1,2)]^T} = \begin{bmatrix}
	   -0.541 & -0.446 & -0.506 & -0.496 & -0.033 & 0.022 & 0.001 & -0.023 & -0.049 & -0.052 \\
	   0.113 & -0.085 & -0.055 & -0.126 & 0.156 & -0.048 & 0.002 & -0.221 & 0.724 & 0.600 \\
	 \end{bmatrix}
\end{align*}
\noindent The variances of the first two sample PCs (again rounded to three decimal places) are 
$\boldsymbol{\hat{\lambda}} = (3.393, 1.521)$. 
The minimum rank 2 reconstruction error for this case (computed as the squared Frobenius norm of the residual matrix) is 597.531 and the Euclidean distance between the first population PC and estimated PC is 0.108.
Results for the SPC, SPC.1se, TPower and rifle methods detailed below on this simple running example can be found in the Supplementary Material.

\subsection{Sparse PCA} \label{sec:sparse_PCA}

As demonstrated by the simple example above, all variables typically make a non-zero contribution to each sample PC even when the data follows a statistical model with sparse population PCs. This property makes interpretation of the PCs challenging, especially when the underlying data is high dimensional. For example, PCA of gene expression data will generate PCs that are linear combinations of thousands of genes and attempting to ascertain the biological relevance of such a large linear combination of genes is a very difficult task.  The challenge of PC interpretation has motivated a large number of approaches for generating approximate PCs that have non-zero loadings for just a small subset of the measured variables. Such sparse PCA techniques include simple components (i.e., PC loading vectors constrained to values from $\{-1,0,1\}$) \citep{vines2000},
methods which compute approximate PCs using cardinality constraints \citep{moghaddam2006spectral, daspremont_direct_2007, Sriperumbudur:2011oq, 10.5555/2567709.2502610, https://doi.org/10.1111/rssb.12291}, methods that use LASSO or elastic net-based penalties \citep{jolliffeSPCA2003, Zou:2006bs, shen2008spca, Witten:2009tg,  doi:10.1080/10618600.2019.1568014}, and methods based on iterative component thresholding \citep{ma2013}. By generating approximate PCs with few non-zero loadings, all of these techniques improve interpretability by associating only a small number of variables with each PC. 
For the comparative evaluation of our proposed approach, we will focus on three current sparse PCA techniques: SPC \citep{Witten:2009tg}, TPower \citep{10.5555/2567709.2502610} and rifle \citep{https://doi.org/10.1111/rssb.12291}. These methods represent both the cardinality constraint and LASSO-penalization approaches and have been shown to provide state-of-the-art performance in both simulation studies and real data analyses. 

The Witten et al. SPC method has an elegant formulation via LASSO-penalized matrix decomposition. Specifically, SPC modifies the standard SVD matrix reconstruction optimization problem to include a LASSO penalty on the PC loadings. Using the notation introduced in Section~\ref{sec:PCA_notation}, the SPC optimization problem for the first PC can be stated as:
\begin{equation}\label{eqn:SPC}
\min_{d,\mathbf{u},\mathbf{v}} ||\mathbf{X} - d \mathbf{u} \mathbf{v}^T||^2_F 
	\text{ subject to } ||\mathbf{u}||^2_2 = 1, ||\mathbf{v}||^2_2 = 1, \sum_{i=1}^n |u_i| < c, d> 0
\end{equation}
\noindent Due to the LASSO penalty on the components of $\mathbf{u}$, this optimization problem generates a sparse version of the first PC loadings vector with the level of sparsity controlled by the parameter $c$. Subsequent sparse PCs can be computed by iteratively applying SPC to the residual matrix generated by subtracting the rank 1 reconstruction $d \mathbf{u} \mathbf{v}^T$ from the original $\mathbf{X}$. Witten et al. also outline a more complex approach for computing multiple PCs that ensures orthogonality. Similar to LASSO-penalized generalized linear models \citep{tibshirani2011lasso}, the penalty parameter $c$ can be specified to achieve a desired level of sparsity in the PC loadings or can be selected via cross-validation to minimize the average out-of-sample matrix reconstruction error.  
Since the true level of sparsity is typically not known, the cross-validation approach is usually employed to determine $c$ and we will use this approach in our comparative evaluation. The need to estimate the sparsity parameter via a technique like cross-validation is not unique to SPC and is a limitation shared by all existing sparse PCA methods.
A common alternative to picking $c$ to minimize the average out-of-sample reconstruction error is to set $c$ to smallest value whose reconstruction error is within 1 standard error of the minimum error. This alternate version, which we will refer to as SPC.1se, generates a more sparse version of the PC loadings at the cost of increased reconstruction error. The SPC and SPC.1se methods are implemented in the PMA R package \citep{Witten:2009tg}.

The Yuan \& Zhang TPower method solves the largest k-sparse eigenvalue problem, which has the following formulation for sparse PCA:
\begin{equation}\label{eqn:tpower}
\max_{\mathbf{v}} \mathbf{v}^T \hat{\boldsymbol{\Sigma}} \mathbf{v}
	\text{ subject to } ||\mathbf{v}||^2_2 = 1, ||\mathbf{v}||_0 \leq k
\end{equation}
\noindent 
The TPower method solves this cardinality constrained optimization problem using a simple truncated version of the power iteration algorithm that truncates the estimated eigenvector on each iteration by setting all loadings to 0 except for the k loadings with the largest absolute values. Similar to the SPC method, the true PC cardinality $k$ is usually unknown and must be estimated using cross-validation. Because an R version of the TPower method is not available, we have provided an implementation in the EESPCA R package along with cross-validation logic to select the sparsity parameter $k$. 


The Tan et al. rifle method solves the largest k-sparse generalized eigenvalue problem, which is identical to \eqref{eqn:tpower} in the context of PCA. For the rifle method, this optimization problem is solved using a two-stage approach that begins with a convex relaxation of \eqref{eqn:tpower} \citep{gao2017} to generate an initial value of the principal eigenvector. This initialization step is followed by a non-convex optimization algorithm similar to the truncated power iteration of TPower that iteratively updates the generalized Rayleigh quotient followed by truncation of the estimated eigenvector to preserve the top $k$ eigenvector loadings with the largest absolute values with all other loadings set to 0. For evaluation of the rifle method, we used the implementation in the rifle R package with cross-validation logic we implemented in the EESPCA package. 


\subsection{Eigenvectors from eigenvalues}

Denton et al. \citep{Denton_2021} recently published a survey of a fascinating association between the squared  elements of the unit length eigenvectors of a Hermitian matrix and the eigenvalues of the full matrix and associated sub-matrices. This identity has a complex history in the mathematical literature with numerous rediscoveries in different disciplines since the earliest known reference in 1834 \citep{JacobiDeBQ}. Importantly in this context, the sample covariance matrix $\hat{\boldsymbol{\Sigma}}$ is Hermitian, which implies that squared PC loadings can be computed as a function of PC variances for the full data set and all of the leave-one-out variable subsets. To briefly restate the results from Denton et al., let $\mathbf{A}$ be a $p \times p$ Hermitian matrix with eigenvalues $\lambda_i(\mathbf{A})$ and unit length eigenvectors $\mathbf{v}_i$ for $i = 1,...,p$. Let the elements of each eigenvector $\mathbf{v}_i$ be denoted $v_{i,j}$ for $j=1,...,p$ and let the squared value of element $j$ be denoted as $|v_{i,j}|^2$.
Let $\mathbf{M}_j$ be the $p-1 \times p-1$ submatrix of $\mathbf{A}$ that results from removing the the $j^{th}$ column and the $j^{th}$ row from $\mathbf{A}$. Denote the eigenvalues of $\mathbf{M}_j$ by $\lambda_k(\mathbf{M}_j$) for $k=1,...,p-1$. Given these values, Denton et al. \citep{Denton_2021} state their key result in Theorem 1:
\vskip 2mm
\noindent \textbf{Theorem 1 (from \citep{Denton_2021}).} \textit{The squared elements of the unit length eigenvectors are related to the eigenvalues and the submatrix eigenvalues:}

\begin{equation} \label{eqn:theorem1}
|v_{i,j}|^2 \prod_{k=1, k \neq i}^{p} (\lambda_i(\mathbf{A}) -\lambda_k(\mathbf{A})) =  \prod_{k=1}^{p-1} (\lambda_i(\mathbf{A}) -\lambda_k(\mathbf{M}_j)) 
\end{equation}

\noindent Which can alternatively be represented as a ratio of the product of eigenvalue differences:

\begin{equation} \label{eqn:lemma2}
|v_{i,j}|^2  = \frac{ \prod_{k=1}^{p-1} (\lambda_i(\mathbf{A}) -\lambda_k(\mathbf{M}_j)) }{ \prod_{k=1, k \neq i}^{p} (\lambda_i (\mathbf{A}) -\lambda_k(\mathbf{A}))} 
\end{equation}

\noindent In the context of PCA, $\mathbf{A}$ can be replaced by the sample covariance matrix $\hat{\boldsymbol{\Sigma}}$ and \eqref{eqn:lemma2} provides a formula for computing squared normed PC loadings from the PC variances of the full matrix and associated sub-matrices.  It is important to note two degenerate scenarios detailed in Denton et al. under which both sides of \eqref{eqn:theorem1} and \eqref{eqn:lemma2} vanish: 1) when $v_{i,j}=0$, $\mathbf{v}_i$ for $\mathbf{A}$ will be an eigenvector for $\mathbf{M}_j$ with the same eigenvalue after deleting the $j^{th}$ coefficient, and 2) when $\mathbf{A}$ has a repeat eigenvalue $\lambda_i(\mathbf{A})$, $\lambda_i(\mathbf{A})$ is also an eigenvalue of $\mathbf{M}_j$. Although the ratio \eqref{eqn:lemma2} is undefined under the second degeneracy scenario (i.e., the denominator is 0), this ratio has the defined value of 0 under the first scenario, which is correct since the eigenvector loading $v_{i,j}$ is 0 in this case. In the context of PCA on experimental data (i.e., $\mathbf{A} = \boldsymbol{\hat{\Sigma}}$), these degenerate scenarios are extremely unlikely for the first $k$ PCs as long as $k$ is less than the rank of $\mathbf{A}$, so we can in practice safely assume that our approximation of \eqref{eqn:lemma2} defined as \eqref{eqn:lemma2_pca_approx} in Section \ref{sec:approx} below is defined for $\boldsymbol{\hat{\Sigma}}$.

In the remainder of this paper, we will detail an approximation of \eqref{eqn:lemma2} relevant to the sparse PCA problem, our proposed EESPCA method (and EESPCA.cv variant) based on this approximation and results from a comparative evaluation of the EESPCA, EESPCA.cv, SPC, SPC.1se, TPower and rifle techniques on simulated data and real data generated using single cell RNA-sequencing \citep{Tanay:2017kq, Wagner:2016lq}. 
An implementation of this method can be found in the EESPCA R package on CRAN. The EESPCA R package also contains a vignette showing results on the simple running example.

\section{Methods}

\subsection{Approximate eigenvector from eigenvalue formulation}\label{sec:approx}

If $\mathbf{A}$ is replaced by the sample covariance matrix $\hat{\boldsymbol{\Sigma}}$, $\hat{\boldsymbol{\Sigma}}_j$ represents the sample covariance sub-matrix with variable $j$ removed, and computation focuses on the loadings for the first PC, \eqref{eqn:lemma2} can be restated as:
\begin{equation} \label{eqn:lemma2_pca}
|v_{1,j}|^2  = \frac{ \prod_{k=1}^{p-1} (\lambda_1(\hat{\boldsymbol{\Sigma}}) -\lambda_k(\hat{\boldsymbol{\Sigma}}_j)) }
	{ \prod_{k=2}^{p} (\lambda_1(\hat{\boldsymbol{\Sigma}}) -\lambda_k(\hat{\boldsymbol{\Sigma}}))} 
\end{equation}

\noindent Although a theoretically interesting result, \eqref{eqn:lemma2_pca} does not have obvious practical utility for PCA since it only generates squared values (i.e., it fails to capture the sign of the loading) and requires computation of the eigenvalues of all of the covariance sub-matrices. However, if we make the approximation that the eigenvalues $\lambda_i(\hat{\boldsymbol{\Sigma}})$ for $i=2,...,p-1$ are equal to the corresponding sub-matrix eigenvalues $\lambda_i(\hat{\boldsymbol{\Sigma}}_j)$, many of the eigenvalue difference terms in \eqref{eqn:lemma2_pca} cancel and we can simplify as follows:
\begin{align*}
|v_{1,j}|^2  &= \frac{ \prod_{k=1}^{p-1} (\lambda_1(\hat{\boldsymbol{\Sigma}}) -\lambda_k(\hat{\boldsymbol{\Sigma}}_j)) }
	{ \prod_{k=2}^{p} (\lambda_1(\hat{\boldsymbol{\Sigma}}) -\lambda_k(\hat{\boldsymbol{\Sigma}}))}  \\
|v_{1,j}|^2  &= \frac{ (\lambda_1(\hat{\boldsymbol{\Sigma}}) -\lambda_1(\hat{\boldsymbol{\Sigma}}_j)) 
	                        (\lambda_1(\hat{\boldsymbol{\Sigma}}) -\lambda_2(\hat{\boldsymbol{\Sigma}}_j)) ...
			       (\lambda_1(\hat{\boldsymbol{\Sigma}}) -\lambda_{p-1}(\hat{\boldsymbol{\Sigma}}_j)) }
		  	     { (\lambda_1(\hat{\boldsymbol{\Sigma}}) -\lambda_2(\hat{\boldsymbol{\Sigma}})) ...
		  	       (\lambda_1(\hat{\boldsymbol{\Sigma}}) -\lambda_{p-1}(\hat{\boldsymbol{\Sigma}}))
			       (\lambda_1(\hat{\boldsymbol{\Sigma}}) -\lambda_p(\hat{\boldsymbol{\Sigma}})) } \\
|\tilde{v}_{1,j}|^2  &= \frac{ \lambda_1(\hat{\boldsymbol{\Sigma}}) -\lambda_1(\hat{\boldsymbol{\Sigma}}_j) }
		  	     {\lambda_1(\hat{\boldsymbol{\Sigma}}) -\lambda_p(\hat{\boldsymbol{\Sigma}}) } 	       
\end{align*}

\noindent If we also assume that $\lambda_p(\hat{\boldsymbol{\Sigma}}) \approx 0$, we can further simplify the approximate squared loadings $|\tilde{v}_{1,j}|^2$ to:
\begin{equation}\label{eqn:lemma2_pca_approx}
|\tilde{v}_{1,j}|^2 = 1 - \frac{\lambda_1(\hat{\boldsymbol{\Sigma}}_j) }
	{ \lambda_1(\hat{\boldsymbol{\Sigma}})}  
\end{equation}

\noindent This approximation has several appealing properties in the context of sparse PCA. 
First, it greatly reduces the computational cost since only the largest eigenvalues of the full covariance matrix and associated sub-matrices are needed, which can be efficiently computed even for very large matrices using the method of power iteration. The computational cost can be further lowered by using the estimated principal eigenvector of the full matrix to initialize the power iteration calculation for the sub-matrices. When this approximation is applied to the population covariance matrix $\boldsymbol{\Sigma}$, it will correctly estimate zero squared loadings since $\lambda_1(\boldsymbol{\Sigma}) = \lambda_1(\boldsymbol{\Sigma}_j)$ for these variables per the first degeneracy scenario outlined above; for variables that have a non-zero loading on the first population PC, the approximation will be less than or equal to the squared value of the true loading. It should also be noted that the second degeneracy case outlined above does not apply to approximation \eqref{eqn:lemma2_pca_approx} since only the terms for the first eigenvalue are retained. For the example introduced in Section~\ref{sec:example}, the squared loadings for the first population PC are:
\begin{align*} 
	 |\mathbf{v}_1|^2 = \begin{bmatrix}
		0.25 & 0.25 & 0.25 & 0.25 & 0 & 0 & 0 & 0 & 0 & 0
	 \end{bmatrix}
\end{align*}

\noindent and the approximate squared loadings computed via \eqref{eqn:lemma2_pca_approx} are:
\begin{align*} 
	 |\mathbf{\tilde{v}}_1|^2 = \begin{bmatrix}
		0.2 & 0.2 & 0.2 & 0.2 & 0 & 0 & 0 & 0 & 0 & 0
	 \end{bmatrix}
\end{align*}

When applied to the sample covariance matrix $\hat{\boldsymbol{\Sigma}}$, the approximate squared loadings will again be less than or equal to the squared PC loadings with the values for variables with a zero loading on the population PC approaching zero as $n \to \infty$ and $n/p \to \infty$ given the consistency of the sample PC loading vectors in this asymptotic regime \citep{johnstone_consistency_2009}. Let $\mathbf{r}_i$ be a vector whose elements $r_{i,j}$ are the ratios of the approximate-to-true absolute loadings of eigenvector $\mathbf{v}_i$:
\begin{equation} \label{eqn:r}
r_{i,j} = \sqrt{|\tilde{v}_{i,j}|^2 / |v_{i,j}|^2}
\end{equation}
Importantly, these ratios for $\mathbf{v}_1$, i.e, $r_{1,j}$, will tend to be larger for variables that have a non-zero loading on the population PC than for variables with a zero population loading. Although we are not able to formally prove this statement, it is informally based on the fact that the eigenvalue difference term retained in the numerator of the approximation, $\lambda_1(\hat{\boldsymbol{\Sigma}}) -\lambda_1(\hat{\boldsymbol{\Sigma}}_j)$, tends to capture a greater proportion of $|v_{1,j}|^2$ for variables with a true non-zero loading on the first population PC than for variables with a zero loading on the first population PC. If variable $j$ only has a non-zero loading on the first population PC and $\lambda_p(\boldsymbol{\Sigma}) \approx 0$, then it follows from PCA consistency \citep{johnstone_consistency_2009} that $r_{1,j} \to 1$ as $n \to \infty$ and $n/p \to \infty$. In particular, the consistency of PC loadings in that asymptotic regime implies consistency of the squared loadings via \eqref{eqn:lemma2} and, when variable $j$ only has a non-zero value on the first population PC and $\lambda_p(\boldsymbol{\Sigma}) \approx 0$, \eqref{eqn:lemma2} reduces to \eqref{eqn:lemma2_pca_approx}, which means that both the numerator and denominator of $r_{1,j}$ converge to the same value.
On the other hand, variables that have a zero loading on the first population PC will have a sample loading that approaches 0 as $n \to \infty$ and $n/p \to \infty$, and, for finite $n$, have a non-zero value based on random contributions spread across the spectrum of eigenvalue difference terms. In this case, the approximation based on the difference term for just the largest eigenvalue, $\lambda_1(\hat{\boldsymbol{\Sigma}}) -\lambda_1(\hat{\boldsymbol{\Sigma}}_j)$, will tend to underestimate the sample loading by a greater degree than for variables with a true non-zero population PC loading. This property of approximation \eqref{eqn:lemma2_pca_approx} provides important information regarding the true sparsity of the PC loadings and plays a key role in the EESPCA method detailed below. For the example data, the ratio (rounded to three decimal places) of approximate to real normed loadings for the first sample PC are:
\begin{align*} 
	 \mathbf{r}_1 = \begin{bmatrix}
	 0.919 & 0.938 & 0.909 & 0.910 & 0.837 & 0.839 & 0.867 & 0.816 & 0.815 & 0.840
	 \end{bmatrix}
\end{align*}

\begin{figure}[h]
\begin{center}
\includegraphics[width=0.8\textwidth]{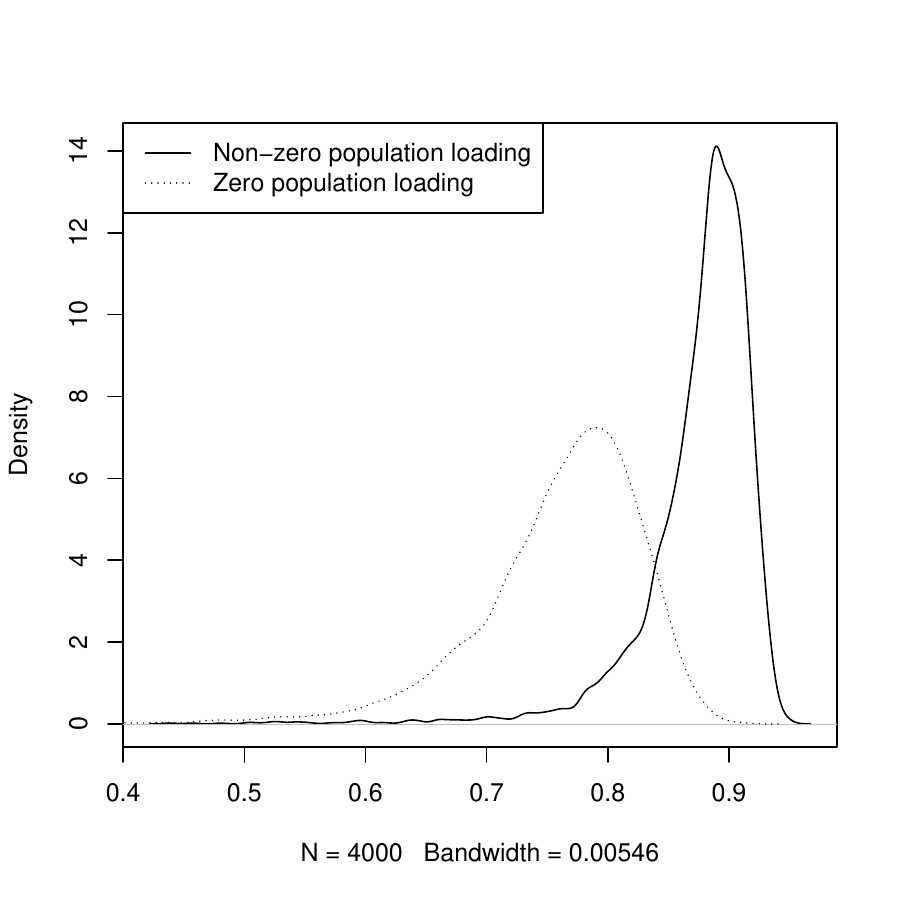}
\end{center}
\caption{Kernel density estimates of the distribution of the ratio of approximate-to-real PC loadings. The ratio distribution for variables that have a non-zero loading on the population PC is shown as a solid line. The ratio distribution for variables that have a zero population loading is shown as a dashed line.}
\label{fig:ratio_distribution}
\end{figure}

Figure~\ref{fig:ratio_distribution} illustrates the relationship between approximate and true squared loadings for first PC as computed on 1,000 data sets simulated according to the model in Section~\ref{sec:example}. As shown in this figure, the ratio for variables with a non-zero loading on the population PC is markedly larger on average than the ratio for variables with a zero population loading.

\clearpage

\subsection{EESPCA}\label{sec:eespca}

\begin{algorithm}\label{alg:eespca}
\caption{Eigenvectors from Eigenvalues Sparse PCA (EESPCA)}
\hspace*{\algorithmicindent} \textbf{Input:} $n \times p$ mean-centered matrix $\mathbf{X}$ and threshold $\alpha$ \\
\hspace*{\algorithmicindent} \textbf{Output:} sparse first PC loadings vector $\mathbf{v}^s_1$ and associated eigenvalue $\lambda^s_1$
\begin{algorithmic}[1]
\State $\hat{\boldsymbol{\Sigma}} = 1/(n-1) \mathbf{X}^T \mathbf{X}$ 
	\Comment{Compute sample covariance matrix}
\State $\{\mathbf{v}_1, \lambda_1\}$ = powerIteration($\hat{\boldsymbol{\Sigma}}$) 
	\Comment{Compute first eigenvector/eigenvalue}
\State $\forall_{j \in 1...p } \lambda_1(\hat{\boldsymbol{\Sigma}}_j) =$ powerIteration($\hat{\boldsymbol{\Sigma}_j}$) 
	\Comment{Compute first submatrix eigenvalues}
\State $\forall_{j \in 1...p } |\tilde{v}_{1,j}|^2  = 1 - \lambda_1(\hat{\boldsymbol{\Sigma}}_j) /  \lambda_1(\hat{\boldsymbol{\Sigma}})$
	\Comment{Approximate first eigenvector using \eqref{eqn:lemma2_pca_approx}}
\State $\forall_{j \in 1...p } r_{1,j} = \sqrt{|\tilde{v}_{1,j}|^2 / |v_{1,j}|^2}$
	\Comment{Compute ratio of eigenvector elements}
\State Compute sparse $\mathbf{v}^s_1$ using $\mathbf{r}_1$:
\begin{algsubstates}
\State $\mathbf{v}^s_1 = \mathbf{r}_1 \odot \mathbf{v}_1$
	\Comment{Scale $\mathbf{v}_1$ by $\mathbf{r}_1$}
\State $\mathbf{v}^s_1 = \mathbf{v}^s_1/||\mathbf{v}^s_1||_2$
	\Comment{Normalize $\mathbf{v}^s_1$ to unit length}
\State Set $v^s_{1,j} = 0$ for all $v^s_{1,j} < \alpha$
	\Comment{Truncate $\mathbf{v}^s_1$ using threshold $\alpha$}
\State $\mathbf{v}^s_1 = \mathbf{v}^s_1/||\mathbf{v}^s_1||_2$
	\Comment{Normalize $\mathbf{v}^s_1$ to unit length}
\end{algsubstates}
\State $\lambda^{s}_1 = (\mathbf{v}^{s}_1)^T \hat{\boldsymbol{\Sigma}} \mathbf{v}^s_1$
	\Comment{Compute sparse first eigenvalue}
\end{algorithmic}
\end{algorithm}

Our proposed EESPCA method, which is based on the eigenvector from eigenvalue approximation \eqref{eqn:lemma2_pca_approx} detailed above, computes a sparse version of the first sample PC for mean-centered $n \times p$ matrix $\mathbf{X}$ using Algorithm 1. We explore two versions of the EESPCA method in the remainder of this paper: a version where $\alpha=1/\sqrt{p}$ (element of a unit length vector that has identical values) and a version where $\alpha$ is selected via cross-validation to minimize the out-of-sample reconstruction error (i.e., the approach used by SPC to select $\lambda$). Both versions of the EESPCA method are implemented in the EESPCA R package, which is available on CRAN. In the reminder of this paper, we will refer to the version where $\alpha=1/\sqrt{p}$ as EESPCA and the version were $\alpha$ is selected by cross-validation as EESPCA.cv. The rationale for using $\alpha=1/\sqrt{p}$ is discussed at greater length in Section \ref{sec:threshold} below. Addition details for each step of the EESPCA algorithm are as follows:

\begin{enumerate}
\item Compute the unbiased sample covariance matrix $\hat{\boldsymbol{\Sigma}} = 1/(n-1) \mathbf{X}^T \mathbf{X}$.
\item Use the power iteration method to compute the principal eigenvector $\mathbf{v}_1$ and associated eigenvalue $\lambda_1$ of $\hat{\boldsymbol{\Sigma}}$.
\item Use the power iteration method to compute the principal eigenvalues of the sub-matrices, $\lambda_1(\hat{\boldsymbol{\Sigma}}_j)$. For efficient calculation, the power iteration method is used with the initial value set to a subsetted version of $\mathbf{v}_1$, i.e., $\mathbf{v}_1$ with the relevant variable removed.
\item Use formula \eqref{eqn:lemma2_pca_approx} to approximate the squared elements of the principal eigenvector $|\mathbf{\tilde{v}_1}|^2$. 
\item Compute the element-wise ratio of approximate-to-real eigenvectors: $r_{1,j} = \sqrt{|\tilde{v}_{1,j}|^2 / |v_{1,j}|^2}$.
\item Use $\mathbf{r}_1$ to compute a sparse version of $\mathbf{v}_1$. As outlined in Section~\ref{sec:approx} above, the entries in $\mathbf{r}_1$ will be less than one and will tend to be larger for variables with a non-zero population loading than for variables with a zero population loading. This property suggests that one could use $\mathbf{r}_1$ to scale $\mathbf{v}_1$ followed by renormalization to generate sample PC loadings that more closely match the population loadings, however, this would not generate sparse loadings since the elements of $\mathbf{r}_1$ are themselves non-zero. To produce a sparse version of $\mathbf{v}_1$, we follow the initial scaling and renormalization 
with a thresholding operation that sets any adjusted loadings less than $\alpha$ to 0 and then renormalize a final time to produce a sparse and unit length loadings vector $\mathbf{v}^s_1$. Specifically:
	\begin{enumerate}
	\item Scale $\mathbf{v}_1$ by $\mathbf{r}_1$, i.e., set $\mathbf{v}^s_1$ to the element-wise product of $\mathbf{v}_1$ and $\mathbf{r}_1$.
	\item Normalize $\mathbf{v}^s_1$ to unit length, i.e., $\mathbf{v}^s_1 = \mathbf{v}^s_1/||\mathbf{v}^s_1||_2$.
	\item Truncate $\mathbf{v}^s_1$ by setting $v^s_{1,j} = 0$ for all $v^s_{1,j} < \alpha$
	\item Normalize $\mathbf{v}^s_1$ again to generate a sparse and unit length eigenvector.
	\end{enumerate}
\item Compute the associated eigenvalue $\lambda^s_1$ as:
\begin{equation}\label{eqn:eespca_lambda}
\lambda^{s}_1 = (\mathbf{v}^{s}_1)^T \hat{\boldsymbol{\Sigma}} \mathbf{v}^s_1
\end{equation}
\end{enumerate}

\noindent To compute multiple sparse PCs, the EESPCA method uses a similar approach to that employed by the SPC method, i.e., it repeatedly applies to the procedure outlined above for calculating the first sparse PC to the residual matrix formed by subtracting the rank 1 reconstruction of X generated using $\mathbf{v}^{s}_1$ from the input $\mathbf{X}$. Note that multiple sparse PCs generated using this recursive approach are not guaranteed to be orthogonal. When applied to the example data described in Section~\ref{sec:example}, the EESPCA method produces the following sparse PC loading vectors (rounded to three decimal places):
\begin{align*} 
	 \mathbf{\hat{V}_{\text{EESPCA}}[,(1,2)]^T} = \begin{bmatrix}
	   0.543 & 0.457 & 0.503 & 0.494 & 0 & 0 & 0 & 0 & 0 & 0 \\
	   0 & 0 & 0 & 0 & 0 & 0 & 0 &  & 0.779 & 0.627 \\
	 \end{bmatrix}
\end{align*}
For this example, the EESPCA.cv version generates identical results since the $\alpha$ selected via 5-fold cross-validation out of 21 potential values equally spaced from $0.75/\sqrt{p}$ to $1.25/\sqrt{p}$ is the default value of $1/\sqrt{p}$.

\subsection{Simulation study design}\label{sec:sim_design}

To explore the comparative performance of the EESPCA, EESPCA.cv, SPC, SPC.1se, TPower and rifle methods, we simulated MVN data for a range of sample sizes, variable dimensionality and covariance structures. Although *.1se variants of EESPCA.cv, TPower and rifle are also possible, results are only shown for SPC.1se since the relative performance pattern is similar. Specifically, we simulated data sets containing $n$ independent samples drawn from a $p$-dimensional MVN data with a zero population mean vector and population covariance matrix $\boldsymbol{\Sigma}$ with all variances set to 1 and covariances set to either 0 or a non-zero $\rho$; 50 data sets were simulated and analyzed for each unique combination of parameter values. The specific $n$, $p$, and $\rho$ values were set according to four different simulation models:
\begin{itemize}
\item \textbf{Basic}: A single distinguishable PC was simulated by setting the covariance between a block of $\beta p$ variables to $\rho$ and to 0 otherwise. According to this covariance model, the first population PC will have equal, non-zero loadings for the first $\beta p$ variables and zero loadings for the remaining $(1-\beta)p$ variables. Simulations were performed for $n$ from 25 to 250, $p$ from 20 to 200, $\rho$ from 0.025 to 0.25, and $\beta$ from 0.025 to 0.25.
\item \textbf{Block covariance}: Three distinguishable PCs were simulated by setting the covariance to $\rho$ for disjoint variable blocks of size $\beta p$, $0.5\beta p$, and $0.25 \beta p$. According to this covariance model, the first population PC will have equal, non-zero loadings for the first $\beta p$ variables and zero loadings for the remaining $(1-\beta)p$ variables, the second population PC will have non-zero loadings for variables in the range $(\beta p, 1.5 \beta p]$, and the third population PC will have non-zero loadings for 
variables in the range $(1.5 \beta p, 1.75 \beta p]$.
Data was simulated for the same $n$, $p$, $\rho$, and $\beta$ ranges used for the basic model.
\item \textbf{High dimension}: Similar to the basic model, a single distinguishable PC was simulated but with larger values of $n$ and $p$ and smaller $\rho$ values. Specifically, data was simulated for $n$ from 50 to 500, $p$ from 100 to 1000, $\rho$ from 0.01 to 0.10, and $\beta$ from 0.025 to 0.25.
\item \textbf{Limited sparsity}: Similar to the basic model, a single distinguishable PC was simulated but with a larger proportion of variables with non-zero loadings on the first population PC.
Specifically, simulations were performed for $n$ from 25 to 250, $p$ from 20 to 200, $\rho$ from 0.025 to 0.25, and $\beta$ from 0.55 to 1.0. The case of $\beta=1.0$ represents a non-sparse scenario with all loadings on the first population PC equal to $1/\sqrt{p}$.
\end{itemize}

The EESPCA, EESPCA.cv, SPC, SPC.1se, TPower and rifle methods were used to compute the first sparse PC of each simulated data set and method performance was quantified according to classification accuracy (i.e., ability of the method to correctly assign zero loadings to variables with a zero population loading), computational speed, out-of-sample rank 1 reconstruction error, and PC estimation error. 
For classification accuracy, the specificity, sensitivity and balanced accuracy of each method were computed. If $v_i$ and $v^s_i$ represent the $i^{th}$ elements of the true eigenvector $\mathbf{v}$ and sparse estimate $\mathbf{v}^s$ and $1()$ is the indicator function, then the sensitivity is defined as $sens = \sum_{i=1}^p 1(v^s_i = 0 \land v_i = 0) / \sum_{i=1}^p 1(v_i = 0)$ (i.e., the proportion of true zero loadings recovered by the sparse PCA method), specificity is defined as $spec = \sum_{i=1}^p 1(v^s_i != 0 \land v_i != 0) / \sum_{i=1}^p 1(v_i != 0)$ (i.e., the proportion of true non-zero loadings recovered by the sparse PCA method), and 
balanced accuracy is defined as $balacc = (sens + spec)/2$.
To measure out-of-sample reconstruction error, the first PC loadings estimated on one simulated data set were used to generate a rank 1 reconstruction of a second data set simulated using identical parameters; reconstruction error was quantified by the squared Frobenius norm of the residual matrix formed by subtracting the rank 1 reconstruction from the simulated matrix.
To measure PC estimation error, the Euclidean distance was computed between the absolute value of the estimated PC loadings and the absolute value of the population PC loadings (absolute values were used to account for potential differences in eigenvector sign).
The SPC and SPC.1se methods were realized using the \textit{SPC()} method in version 1.2.1 of the PMA R package \citep{Witten:2009tg} with the optimal penalty parameter computed using the \textit{SPC.cv()} method with \textit{nfolds=5}, \textit{niter=10} and \textit{sumabsv=seq(1, sqrt(p), len=20)}. We created an R implementation of the TPower algorithm, which is available in the EESPCA R package via the function \textit{tpower()}. The initial eigenvector for the TPower algorithm was computed using non-truncated power iteration. To determine the optimal cardinality parameter $k$ for TPower, we implemented a cross-validation method based on the \textit{SPC.cv()} function. This cross-validation method is available in the EESPCA R package via the function \textit{tpowerPCACV()} and was executed for the simulation studies using \textit{nfolds=5} and \textit{k.value=round(seq(1, p, len=20))}.
The rifle method was realized using the \textit{rifle()} function in version 1.0 of the rifle R package. The initial eigenvector for the rifle method was computed using the suggested \textit{initial.convex()} function from the rifle R package using $K=1$ and $lambda=sqrt(log(p)/n)$ (this is implemented in the EESPCA R package using the convenience function \textit{rifleInit()}). Similar to the TPower method, the optimal cardinality parameter $k$ for rifle was computed using the Witten et al. cross-validation approach as implemented by the \textit{riflePCACV()} function in the EESPCA R package with \textit{nfolds=5} and \textit{k.value=round(seq(1, p, len=20))}. For the EESPCA.cv method, the optimal threshold was selected using the Witten et al. cross-validation approach as implemented by the \textit{eespcaCV()} function the EESPCA R package with \textit{nfolds=5} and 
\textit{sparse.threshold.values=seq(from=0.75/sqrt(p), to=1.25/sqrt(p), length.out=21)}.

\subsection{Real data analysis design}\label{sec:real_design}

To explore the performance of the EESPCA, EESPCA.cv, SPC, SPC.1se, TPower and rifle methods on real data, we analyzed a single cell RNA-sequencing (scRNA-seq) data set generated on 2.7k human peripheral blood mononuclear (PBMC) cells. This scRNA-seq data set is freely available from 10x Genomics and is used in the Guided Clustering Tutorial \citep{SeuratCluteringTutorial} for the Seurat single cell framework \citep{Stuart:2019yq}. 
Preprocessing, quality control (QC) and normalization of the PBMC data set followed the same processing steps used in the Seurat Guided Clustering Tutorial. Specifically, the Seurat log-normalization method was used followed by application of the \textit{vst} method for decomposing technical and biological variance. Seurat log-normalization divides the unique molecular identifier (UMI) counts for each gene in a specific cell by the sum of the UMI counts for all genes measured in the cell and multiplies this ratio by the scale factor $1\times10^6$. The normalized scRNA-seq values are then generated by taking the natural log of this relative value plus 1. This technique generates normalized data whose non-zero values can be approximated by a log-normal distribution. The Seurat \textit{vst} method fits a non-linear trend to the log scale variance/mean relationship. This estimated trend models the expected technical variance based on mean gene expression; observed variance values above this expected trend reflect biological variance.
Preprocessing and QC of the PBMC data yielded normalized counts for 14,497 genes and 2,638 cells. Immune cell types were assigned using the same procedure detailed in the Seurat Guided Clustering Tutorial.

The 1,000 genes with the largest estimated biological variance according to the \textit{vst} method were used as input for standard PCA, EESPCA, EESPCA.cv, SPC, SPC.1se, TPower and rifle. Only the first two sparse PCs were generated using the five sparse PCA methods, which were executed using the same parameter settings as detailed in Section~\ref{sec:sim_design}. To capture out-of-sample reconstruction error, the scRNA-seq data set was randomly split in half 10 times and, for each split, PCs were estimated using each method on one half and the squared Frobenius norm of the residual matrix for the first two PCs was computed onthe other half of the data. Gene Ontology (GO) \citep{Gene-Ontology-Consortium:2010nx} enrichment analysis was used to interpret the normal and sparse PCs. Specifically, the \textit{goana()} method \citep{Young:2010aa} in the \textit{limma} R package \citep{Ritchie:2015yu} was used to determine the statistical enrichment of GO Biological Process Ontology terms among genes with either positive or negative PC loadings (GO annotations were obtained using version 3.11.4 of the GO.db Bioconductor R package \citep{Carlson:2020tu}). The \textit{goana()} method performs this enrichment analysis using a Fisher's exact test on the $2 \times 2$ contingency table that categorizes the 1,000 high variance genes according to whether they belong to the target GO term and whether than have a positive (or negative) loading on the target PC. False discovery rate (FDR) values were computed using the Benjamini and Hochberg method \citep{benjamini1995controlling} separately for each PC and loading direction for the family of test corresponding to all GO Biological Process terms.

\section{Results and discussion}\label{sec:results}

The relative performance of the evaluated methods on data simulated according to the basic model is detailed in Sections \ref{sec:classification}-\ref{sec:PC_error} below. 
The motivation for using $\alpha = 1/\sqrt{p}$ for the EESPCA method is outlined in Section \ref{sec:threshold} and results for the scRNA-seq example are contained in Section \ref{sec:real_results}.

\subsection{Classification performance}\label{sec:classification}

\begin{figure}[t]
\begin{center}
\includegraphics[width=0.9\textwidth]{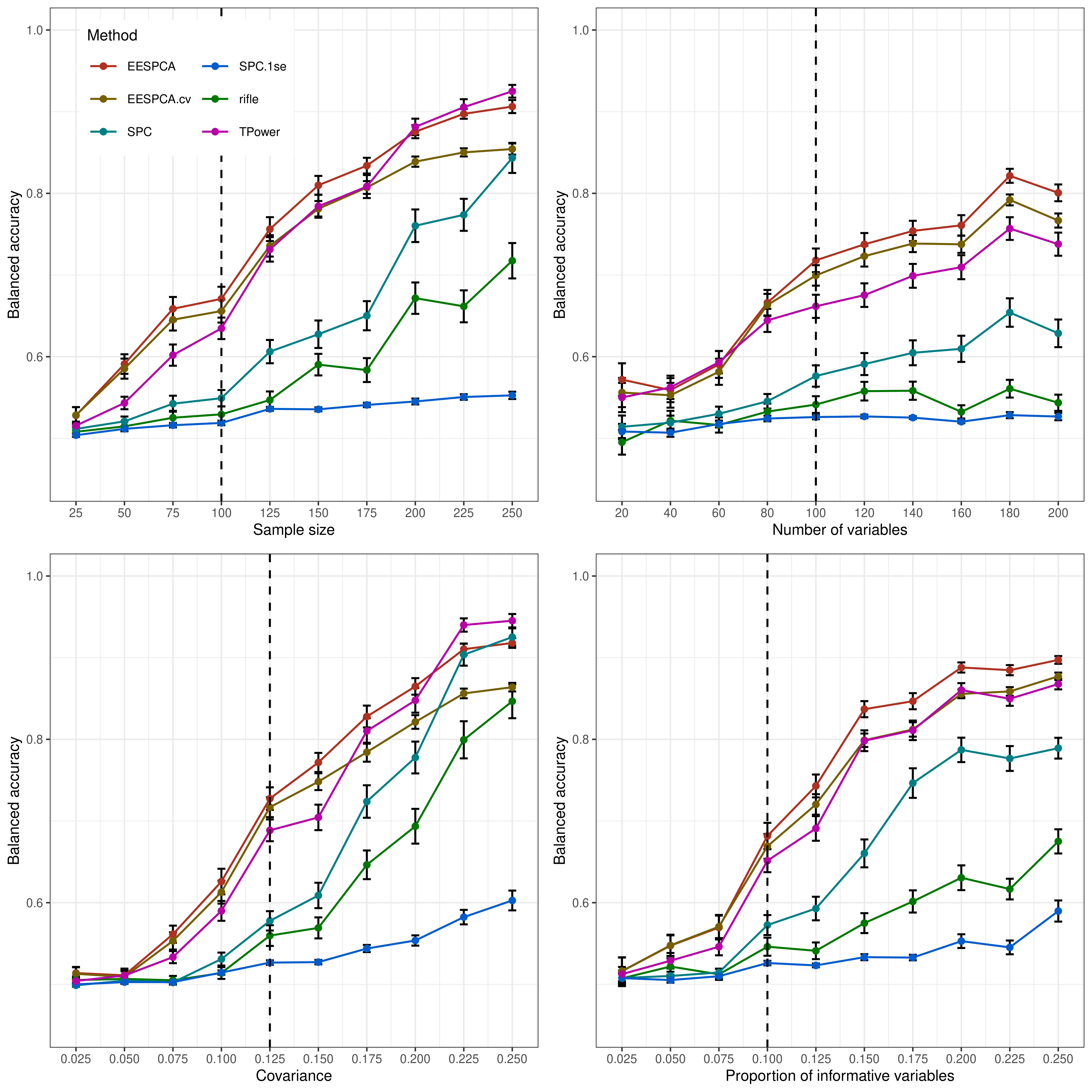}
\end{center}
\caption{Classification performance of EESPCA, EESPCA.cv, SPC, SPC.1se, TPower and rifle on data simulated using the basic model detailed in Section~\ref{sec:sim_design}. Each panel illustrates the relationship between the balanced accuracy of the methods (i.e., the ability to correctly assign 0 loadings to variables with a 0 population loading) and one of the simulation parameters. The vertical dotted lines mark the default parameter value used in the other panels. Error bars represent $\pm 1$ SE.}
\label{fig:accuracy}
\end{figure}

\begin{figure}[t]
\begin{center}
\includegraphics[width=0.9\textwidth]{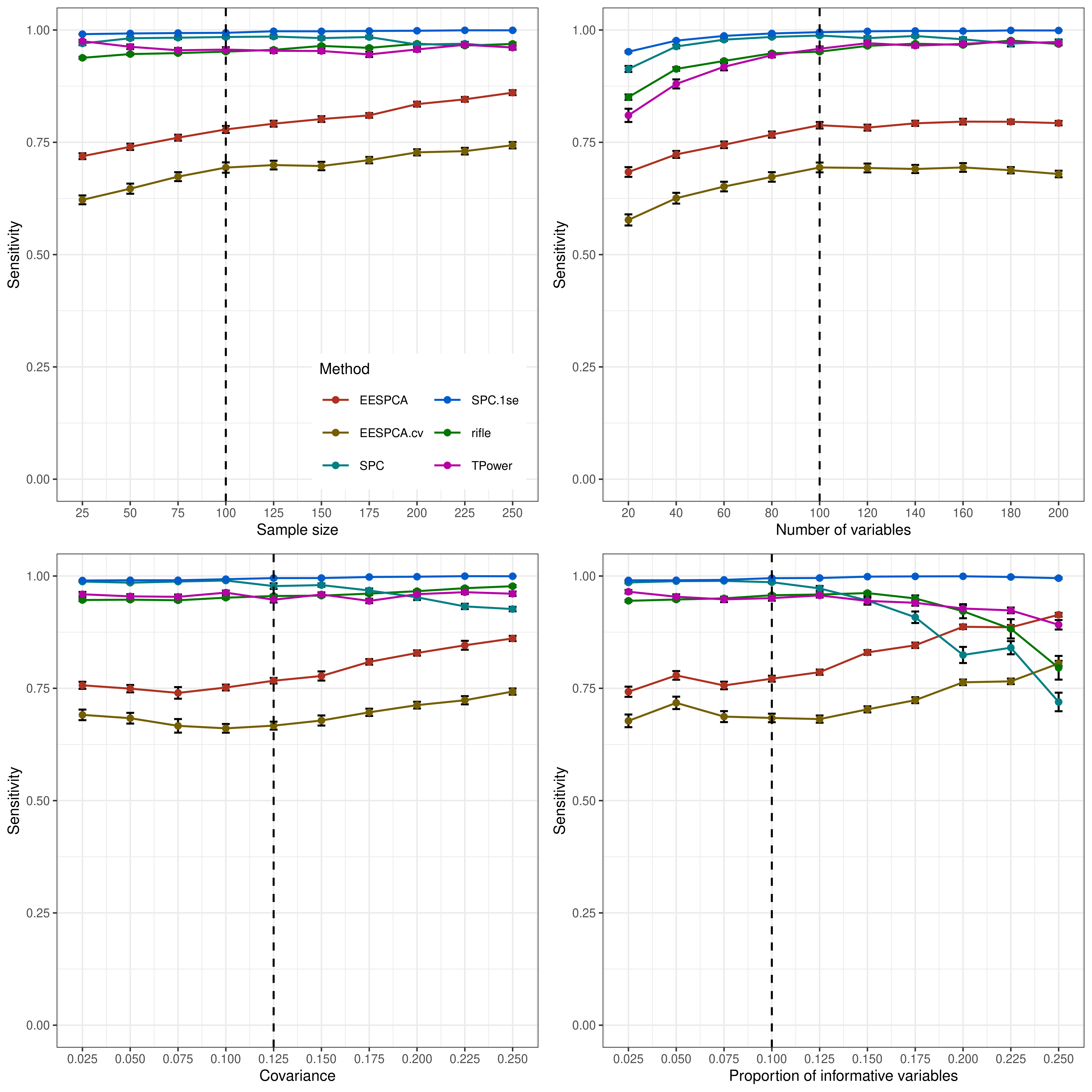}
\end{center}
\caption{Sensitivity of EESPCA, EESPCA.cv, SPC, SPC.1se, TPower and rifle on data simulated using the basic model detailed in Section~\ref{sec:sim_design}.  Each panel illustrates the relationship between sensitivity (i.e., the proportion of variables with a zero population loading that are given a zero loading in the estimated sparse PC) and one of the simulation parameters. The vertical dotted lines mark the default parameter value used in the other panels. Error bars represent $\pm 1$ SE.}
\label{fig:sensitivity}
\end{figure}

\begin{figure}[t]
\begin{center}
\includegraphics[width=0.9\textwidth]{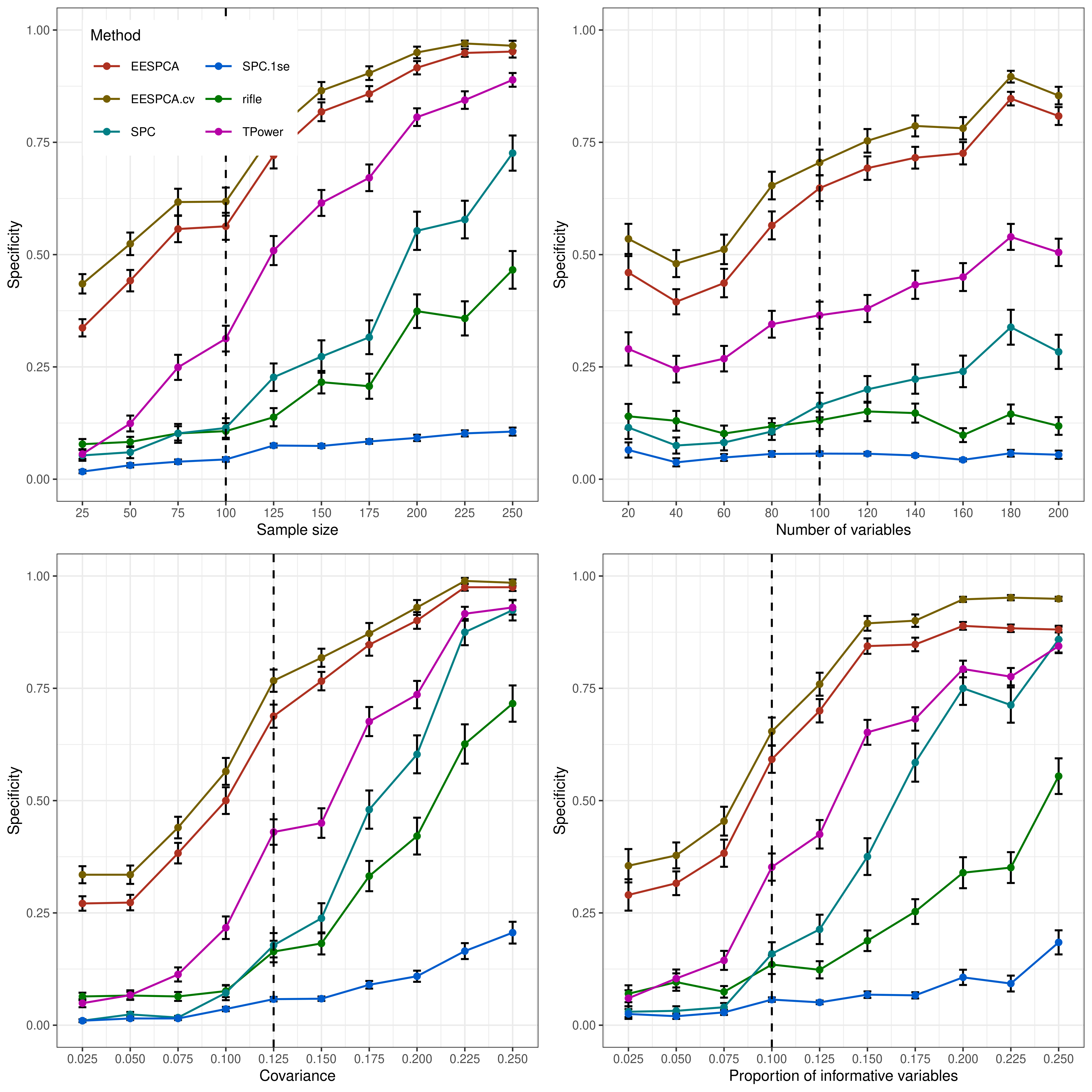}
\end{center}
\caption{Specificity of EESPCA, EESPCA.cv, SPC, SPC.1se, TPower and rifle on data simulated using the basic model detailed in Section~\ref{sec:sim_design}.  Each panel illustrates the relationship between specificity (i.e., the proportion of variables with a non-zero population loading that are given a non-zero loading in the estimated sparse PC) and one of the simulation parameters. The vertical dotted lines mark the default parameter value used in the other panels. Error bars represent $\pm 1$ SE.}
\label{fig:specificity}
\end{figure}

As illustrated in Figure~\ref{fig:accuracy}, the EESPCA method has superior classification performance (as measured by balanced accuracy) relative to the other methods across almost the full range of parameter values explored for the basic model. Performance of the EESPCA.cv and TPower methods was close to EESPCA with TPower having slightly better accuracy at large $n$ and large $\rho$. As expected, the accuracy of all methods tends to increase as either $n$,  $\rho$ or $\beta$ increase; improved performance at larger $p$ with $n$ fixed was unexpected. 
Performance of the SPC and rifle methods approaches the EESPCA, EESPCA.cv and TPower methods at higher parameter values. The corresponding sensitivity (i.e., ability to assign zero loadings for variables with a zero population loading) and specificity (i.e., ability to assign a non-zero loading to variables with a non-zero population loading) values, shown in Figures~\ref{fig:sensitivity} and~\ref{fig:specificity}, provide more insight into the relative performance of the methods. Looking at the sensitivity plots in Figure~\ref{fig:sensitivity}, the SPC.1se method had almost perfect sensitivity, which is consistent with the fact that the SPC.1se method generates a much sparser solution than the SPC variant. Sensitivity of the SPC, rifle and TPower methods was also very high. By contrast, the EESPCA and EESPCA.cv methods have lower sensitivity for this simulation model, although the sensitivity improves as sparsity decreases, i.e., a larger proportion of informative variables. 
As shown in Figure~\ref{fig:specificity}, EESPCA and EESPCA.cv methods have the best specificity across the range of simulated parameter values. As expected, the specificity for all evaluated methods increases as the parameter values increase. Relative to the other methods, SPC.1se has much lower specificity, which is expected given the sparser solution generated by this approach. 

Classification performance for the block covariance, high dimension and limited sparsity models is contained in Figures S1-S3, S7-S9, and S13-S15 in the Supplementary Results. Relative classification performance of the evaluated methods for both the block covariance model and high dimension model is very similar to the performance for the basic model, i.e., EESPCA achieves the best balanced accuracy across almost all evaluated parameter values with EESPCA.cv and TPower just slightly worse. For the limited sparsity model, however, EESPCA.cv achieves the best balanced accuracy with SPC.1se a close second, EESPCA and TPower significantly lower, and rifle and SPC at the bottom. In this scenario, the poor relative performance of the EESPCA model is due to an overly sparse model (i.e., the $1/\sqrt{p}$ threshold is too high resulting in poor specificity).

\clearpage

\subsection{Computational speed}\label{sec:speed}

\begin{figure}[h]
\begin{center}
\includegraphics[width=0.9\textwidth]{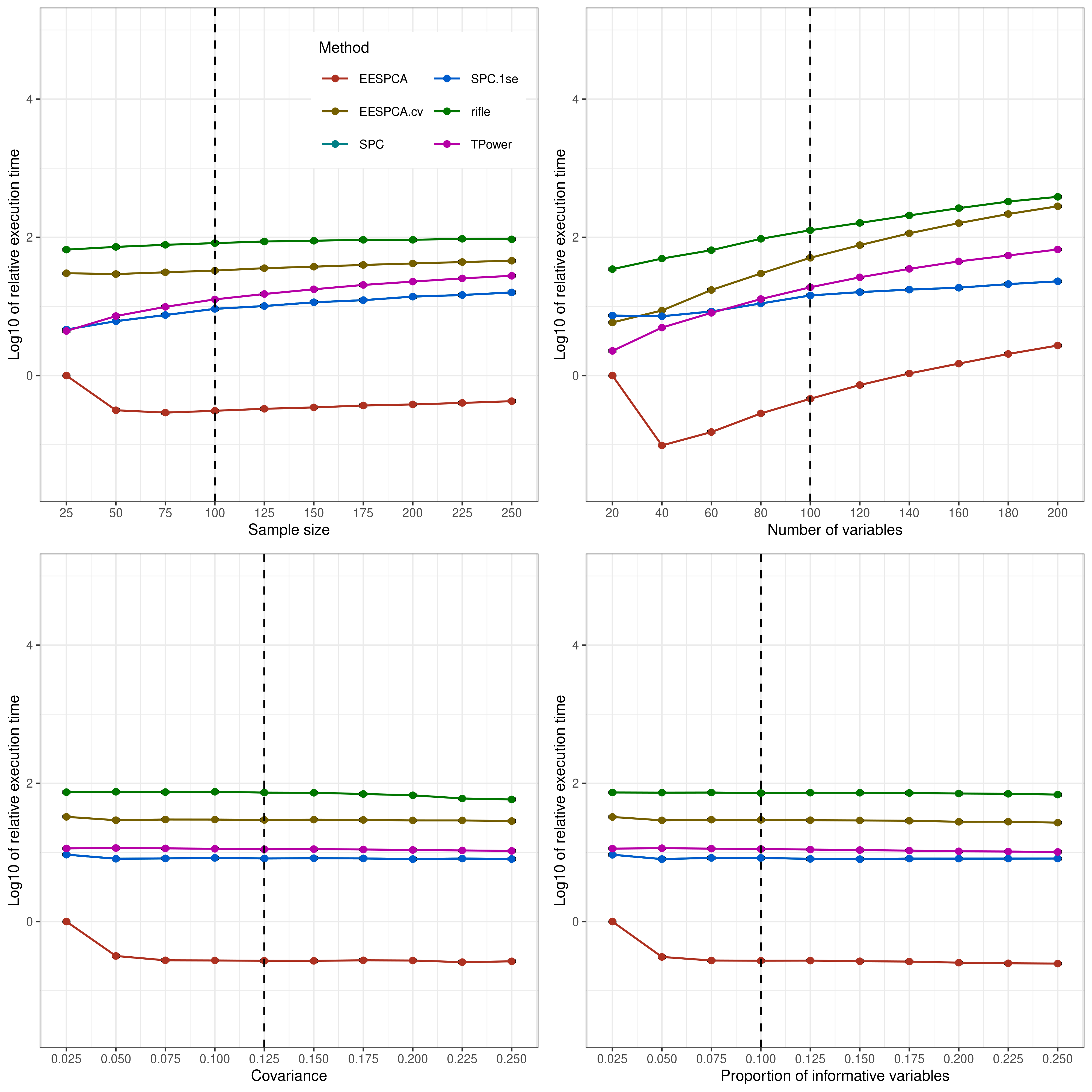}
\end{center}
\caption{Relative computational speed of EESPCA, EESPCA.cv, SPC, SPC.1se, TPower and rifle on data simulated using the basic model detailed in Section~\ref{sec:sim_design}. Each panel illustrates the relationship between the relative computational speed of each method (i.e., log$_{10}$ ratio of computational cost of the method and the cost of the EESPCA method on data simulated using the smallest parameter value shown on the x-axis) and one of the simulation parameters. The SPC and SPC.1se lines overlap given the identical computational times. The vertical dotted lines mark the default parameter value used in the other panels. Error bars represent $\pm 1$ SE.}
\label{fig:speed}
\end{figure}

Figure~\ref{fig:speed} illustrates the relative computational cost of the evaluated methods for the basic simulation model. Because the SPC.1se and SPC methods have equivalent computational cost, the lines overlap and only the SPC.1se values are visible. As shown in this figure, the execution times of the EESPCA.cv, SPC, SPC.1se, rifle and TPower methods are between one and two-orders-of-magnitude larger than the EESPCA method across all tested parameter values. 
 The dramatic differences in computational cost are primarily driven by the required cross-validation-based selection of sparsity constraints for these methods. As expected, computational speed is fairly insensitive to changes in $\rho$ or $\beta$ for all methods and tends to increase as either $n$ or $p$ are increased.

Computational speed for the block covariance, high dimension and limited sparsity models is contained in Figures S4, S10, S16 in the Supplementary Results. The relative computational cost of the evaluated methods for both the block covariance model and limited sparsity model is very similar to the cost measured on the basic model, i.e., the computational cost of EESPCA is between one and two-orders-of-magnitude better than other methods. For the high dimension model, however, the SPC and SPC.1se methods have the best computational cost at large $p$ with fixed $n$. This is due to the fact that the penalized matrix decomposition technique used SPC and SPC.1se can in effect solve the eigenvalue problem for the smaller $n \times n$ matrix $\mathbf{X} \mathbf{X}^T$ rather than for the $p \times p$ matrix $\mathbf{X}^T \mathbf{X}$. 
\clearpage

\subsection{Reconstruction error}\label{sec:recon_error}

\begin{figure}[h]
\begin{center}
\includegraphics[width=0.9\textwidth]{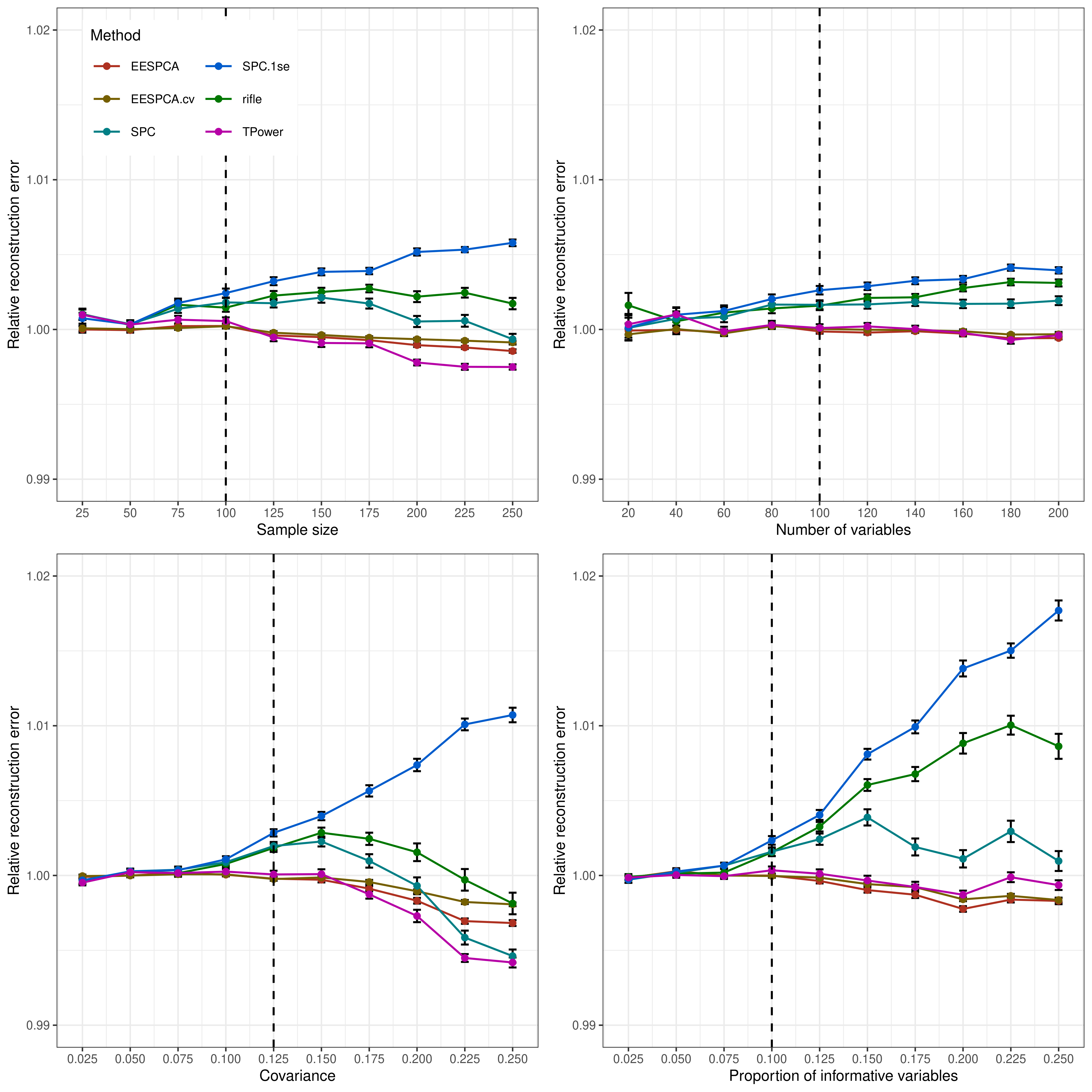}
\end{center}
\caption{Rank 1 out-of-sample reconstruction error of EESPCA, EESPCA.cv, SPC, SPC.1se, TPower and rifle on data simulated using the basic model detailed in Section~\ref{sec:sim_design}. Each panel illustrates the relationship between the ratio of the rank 1 out-of-sample reconstruction error (i.e., squared Frobenius norm of the residual matrix formed by subtracting the rank 1 reconstruction matrix from the original matrix) for each method to the error achieved by standard PCA. The vertical dotted lines mark the default parameter value used in the other panels. Error bars represent $\pm 1$ SE.}
\label{fig:error}
\end{figure}

Figure~\ref{fig:error} shows the relative rank 1 out-of-sample reconstruction error of the evaluated sparse methods as compared to standard PCA for the basic simulation model. The reconstruction error for the EESPCA, EESPCA.cv and TPower methods was similar and, for higher parameter values, lower than standard PCA. Compared to these methods, the reconstruction error for rifle, SPC and SPC.1se was markedly higher. As expected given the more sparse solution, the SPC.1se had the worst reconstruction error of all evaluated methods.

Reconstruction error for the block covariance, high dimension and limited sparsity models is contained in Figures S5, S11, S17 in the Supplementary Results. The relative out-of-sample reconstruction error of the evaluated methods for both the block covariance model and high dimension model is very similar to the reconstruction error measured on the basic model. For the limited sparsity model, however, the EESPCA method has a dramatically larger reconstruction error than the other methods with SPC and TPower demonstrating the best performance. Although the reconstruction error for the EESPCA.cv method in this case is much lower than the error for EESPCA, it is larger than SPC, rifle and TPower. This pattern is in contrast to classification accuracy for the limited sparsity model, where EESPCA.cv had the best performance.
\clearpage

\subsection{PC estimation error}\label{sec:PC_error}

\begin{figure}[h]
\begin{center}
\includegraphics[width=0.9\textwidth]{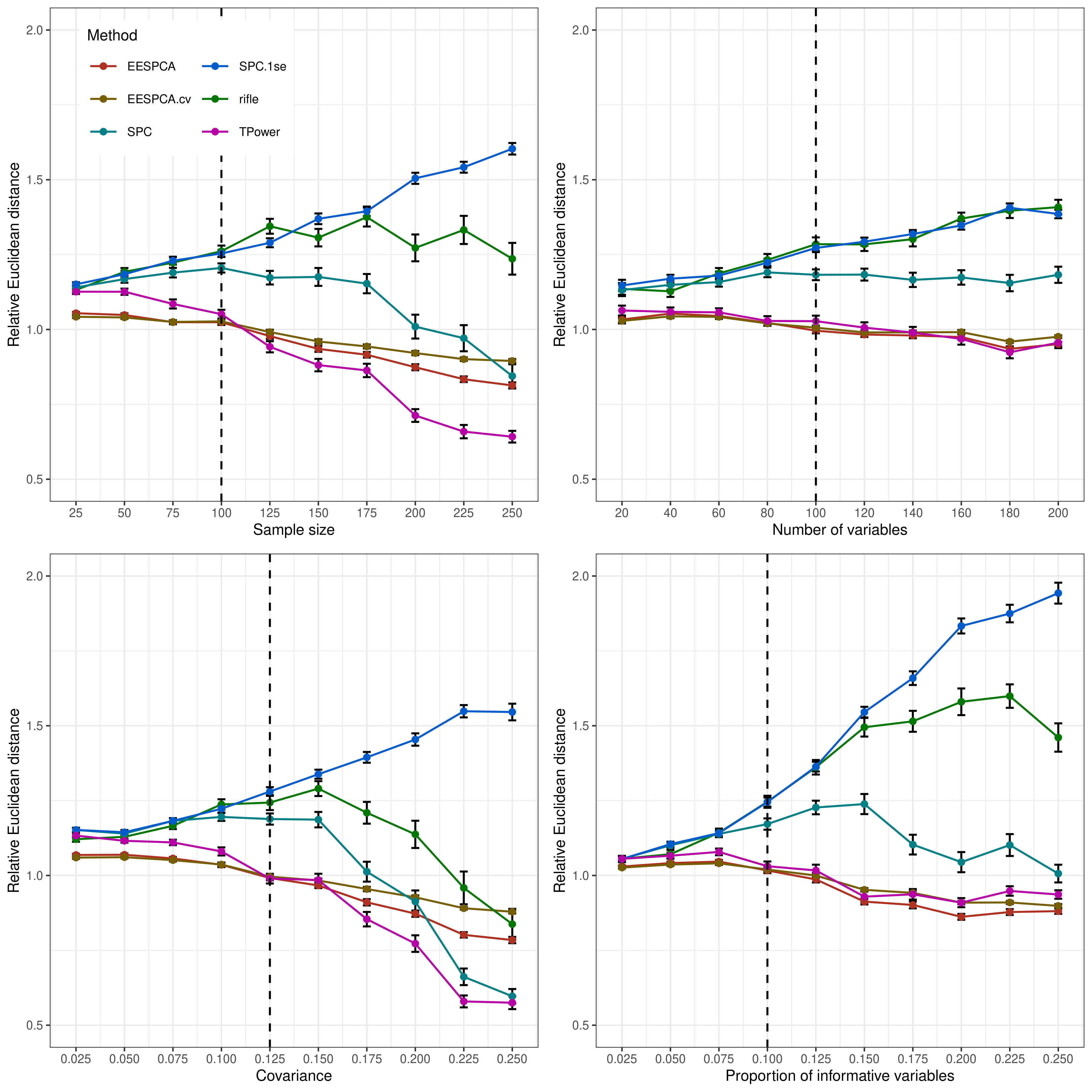}
\end{center}
\caption{Euclidean distance between the estimated and population loadings of the first PC for the basic model
detailed in Section~\ref{sec:sim_design}. Each panel illustrates the relationship between the ratio of the distance for each method to the distance achieved by standard PCA. The vertical dotted lines mark the default parameter value used in the other panels. Error bars represent $\pm 1$ SE.}
\label{fig:L2}
\end{figure}

Figure~\ref{fig:L2} shows the relative PC estimation error of the evaluated sparse methods as compared standard PCA for the basic model. The plotted estimation error is quantified by the Euclidean distance between the estimated PC loadings and the population loadings of the first PC.
As expected, the relative PC estimation error follows a similar trend to the relative out-of-sample reconstruction error.

PC estimation error for the block covariance, high dimension and limited sparsity models is contained in Figures S6, S12, S18 in the Supplementary Results. The relative PC estimation error of the evaluated methods for both the block covariance model and high dimension model is very similar to the PC estimation error measured on the basic model. For the limited sparsity model, however, the EESPCA method has a dramatically larger estimation error than the other methods with SPC and TPower demonstrating the best performance. As expected, the results for PC estimation error on these models are similar to those found for out-of-sample reconstruction error.
\clearpage

\subsection{Rationale for $\alpha=1/\sqrt{p}$}\label{sec:threshold}

Although we are unable to formally prove the optimality of $\alpha=1/\sqrt{p}$, this choice for the default threshold can be motivated by both heuristic arguments and simulation results.
For a unit length eigenvector of length $p$ with identical elements, all elements have the same absolute value of $1/\sqrt{p}$. Importantly, this form of eigenvector occurs as the loadings for the first population PC in the case where all population variances are equal and all covariances are equal to a non-zero $\rho$. In the context of sparse PCA, such a population covariance matrix can be considered as a null case, i.e., a covariance matrix with no internal substructure and first population PC loadings that diverge, in a qualitative sense, as far as possible from a sparse eigenvector. When sparsity does exist, some of the population PC loadings will equal 0, forcing the non-zero elements to be greater than $1/\sqrt{p}$ for a unit length eigenvector. 
For estimation of sample PCs when consistency holds (i.e., $p/n \to 0$ \citep{johnstone_consistency_2009}), the expected absolute value of the loadings will converge in probability to $1/\sqrt{p}$ in the null scenario and to 0 or values above $1/\sqrt{p}$ in the sparse scenario.
These properties of the population and sample PC loadings in the null and sparse cases motivate our use of $1/\sqrt{p}$ as a default threshold for the EESPCA method. Informally, when a sparse population PC is assumed, the non-sparse elements can be expected to have sample values above $1/\sqrt{p}$ and the sparse elements can be expected to have sample values below $1/\sqrt{p}$. The EESPCA method attempts to magnify these deviations away from $1/\sqrt{p}$ by scaling the sample PC loadings by the ratio $\mathbf{r}_1$ \eqref{eqn:r} of approximate-to-real loadings. Because this ratio tends to be larger for non-sparse elements than for sparse elements, the non-sparse elements will be shifted further above $1/\sqrt{p}$ and the sparse elements further below $1/\sqrt{p}$. 

Simulation studies also provide empirical support for using $1/\sqrt{p}$ as a default threshold when significant sparsity can be assumed for the population PC loadings. As shown in Sections \ref{sec:classification}, \ref{sec:recon_error} and \ref{sec:PC_error} and the Supplementary Results, the EESPCA variant performs as well as or better than the EESPCA.cv variant across a wide range of simulation models (i.e., the basic, block covariance and high dimension models detailed in Section \ref{sec:sim_design}) in terms of sparsity estimation (Figures \ref{fig:accuracy}, S1, and S7), out-of-sample reconstruction error (Figures \ref{fig:error}, S5 and S11) and PC estimation error (Figures \ref{fig:L2}, S6, and S12). For these simulation models, the $1/\sqrt{p}$ threshold generates a more sparse solution on average than the threshold selected via cross-validation (i.e., the sensitivity of EESPCA is uniformly larger than the sensitivity of EESPCA.cv as shown in Figures \ref{fig:sensitivity}, S2 and S8 with the reverse holding for specificity as shown in Figures \ref{fig:specificity}, S3 and S9). However, as demonstrated by the results for the limited sparsity model (Figures S13-S18), the performance of the $1/\sqrt{p}$ threshold does not hold when the population sparsity is limited. In this limited sparsity scenario, the $1/\sqrt{p}$ default is too large and the threshold should instead be selected via cross-validation to minimize out-of-sample reconstruction error.

\subsection{Real data analysis}\label{sec:real_results}

\begin{figure}[h]
\begin{center}
\includegraphics[width=0.9\textwidth]{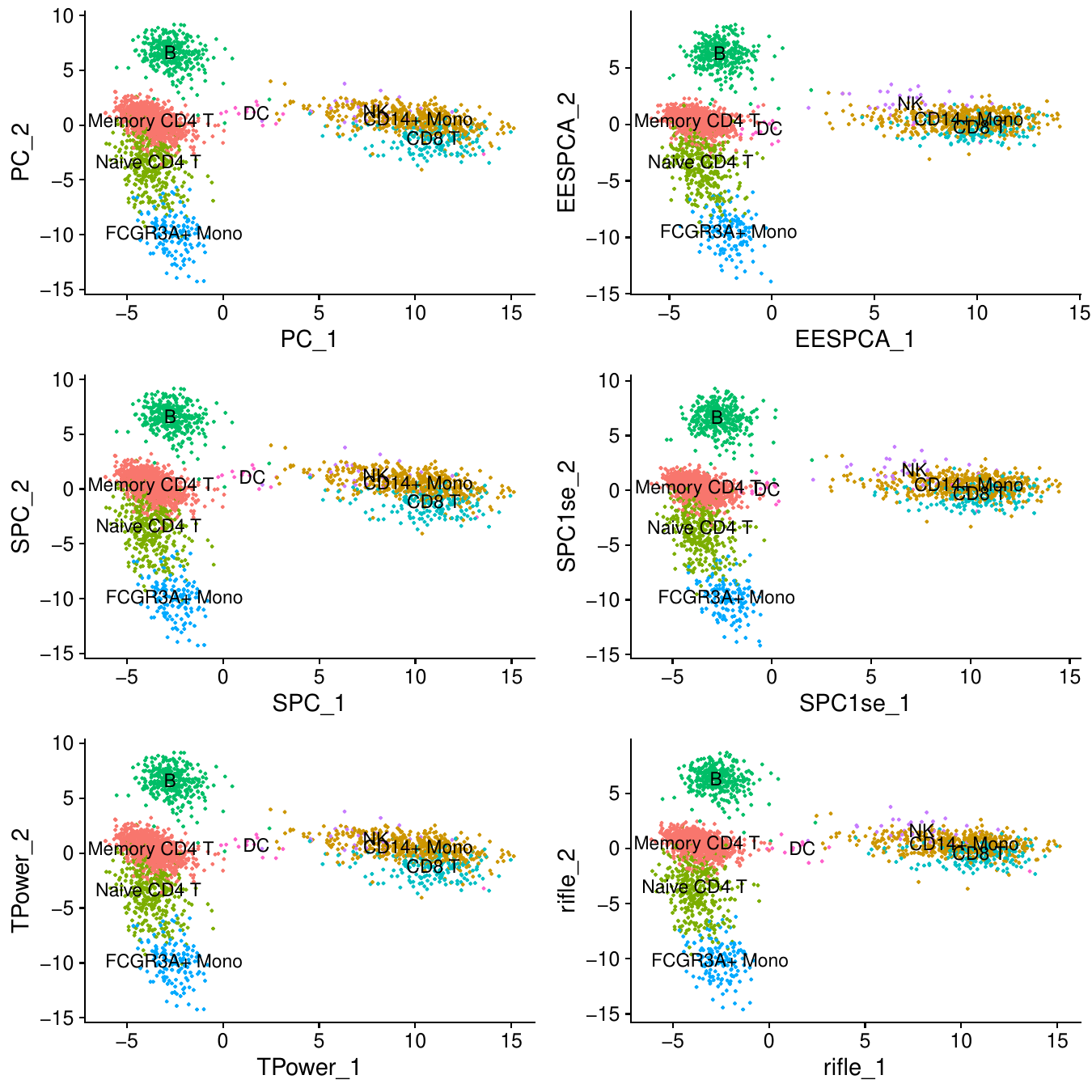}
\end{center}
\caption{Projections of the PBMC 2k single cell RNA-sequencing data onto the first two PC computed using standard PCA or the first two sparse PCs computed using the EESPCA, SPC, SPC.1se, TPower and rifle methods (EESPCA.cv results are very similar to the EESPCA results so have been omitted to enable a 3-by-2 panel plot). Each point represents a single cell is colored and labeled according to immune cell type as detailed in the Seurat Guided Clustering Tutorial \cite{SeuratCluteringTutorial}.}
\label{fig:pbmc}
\end{figure}

As detailed in Section~\ref{sec:real_design}, standard PCA, EESPCA, EESPCA.cv, SPC, SPC.1se, TPower and rifle were applied to a scRNA-seq data set that captured the expression values of the 1,000 genes with the largest estimated biological variance for 2,638 individual human immune cells. As illustrated in Figure~\ref{fig:pbmc}, all methods produce a similar projection of these cells onto the first two standard or sparse PCs (the EESPCA.cv projection is very similar to the EESPCA projection so has been omitted to allow for a 3-by-2 panel plot). Similar to the simulation results, the EESPCA method was close to 2 orders-of-magnitude faster than SPC or SPC.1se (20.87 seconds vs. 25.92 minutes on a standard laptop). The TPower and EESPCA.cv methods were markedly slower, taking 57.12 and 54.07 minutes respectively. Because the rifle method is more than an order of magnitude slower than SPC, cross-validation was not performed for this data set and the optimal $k$ values computed for TPower were used instead; even without cross-validation, execution of rifle still took 7.08 minutes. Qualitatively, all of the methods generate a very similar projection with the first PC separating cytotoxic from non-cytotoxic cells and the second PC capturing phenotypic differences among the non-cytotoxic populations.

Table~\ref{table:pbmc_results} lists various statistics for first two normal or sparse PCs computed by each of these methods on the PBMC scRNA-seq data. Standard PCA assigned a non-zero loading to all genes on both PCs and GO enrichment analysis based on these loadings only produced significant findings for genes associated with PC 1 (44 significant terms for genes with a positive sign).
The SPC method generated very little sparsity on this data set with GO enrichment results similar to those generated by standard PCA. As expected, the SPC.1se method produced sparser PCs than the SPC method, which increased the number of significant GO terms for PC 1 and allowed 47 GO terms to reach significance for PC 2 (33 for the positive direction and 14 for the negative direction). TPower produced few zero loadings for the first PC so had enrichment results similar to standard PCA. For PC 2, TPower generated some sparsity with 3 significant GO terms for the positive direction. Because rifle was executed with the same cardinality constraints as TPower, it also produced similar results for PC 1 and only moderate sparsity for PC 2 with similar numbers of significant GO terms. The EESPC method generated the most sparse solution on this scRNA-seq data, which enabled GO enrichment analysis to produce significant results for three of the four test cases.  All of the sparse PCA methods had average out-of-sample reconstruction errors less than one percent larger than PCA with the SPC error slightly below the error produced by PCA. As expected, methods that produced sparser estimates had a larger out-of-sample reconstruction error. Table~\ref{table:go_results} below lists the five most significant GO terms for each test case for PCA, EESPCA, SPC and TPower (equivalent results for EESPCA.cv, SPC.1se and rifle can be found in the Supplementary Material). Importantly, these results mirror the expected interpretation of the first two PCs. These results also highlight a well known challenge with hierarchical gene set collections like the Gene Ontology, namely that the ranked result list will contain sequences of biologically similar and highly overlapping gene sets. 

\begin{table}
\centering
 \begin{tabular}{l  | c | c | c | c | c | c | c | c | c}
   &  \multicolumn{4}{c|}{PC 1} & \multicolumn{4}{c|}{PC 2} & Relative \\
   & \multicolumn{2}{c|}{Positive} & \multicolumn{2}{c|}{Negative} & \multicolumn{2}{c|}{Positive} & \multicolumn{2}{c|}{Negative} & reconstruction \\
   & genes & sig. GO & genes & sig. GO & genes & sig. GO & genes & sig. GO & error \\
  \hline
PCA & 558 & 44 & 442 & 0 & 316 & 0 & 684 & 0 & 1 (0) \\
  \hline  
EESPCA & 134 & 104 & 12 & 0 & 35 & 38 & 83 & 3 & 1.004 (0.0005) \\
  \hline   
EESPCA.cv & 166 & 89 & 22 & 0 & 41 & 33 & 129 & 14 & 1.003 (0.0004) \\
  \hline     
SPC & 535 & 37 & 418 & 0 & 316 & 0 & 684 & 0 & 0.996 (0.003) \\
  \hline   
SPC.1se & 193 & 84 & 36 & 0 & 91 & 21 & 394 & 1 & 1.003 (0.0002)\\
  \hline   
TPower & 485 & 38 & 357 & 0 & 226 & 3 & 616 & 0 & 1.0002 (0.0003) \\
  \hline   
rifle & 485 & 38 & 357 & 0 & 268 & 2 & 574 & 2 & 1.002 (0.002) \\
  \end{tabular}
  \caption{Results from standard PCA, EESPC, EESPCA.cv, SPC, SPC.1se, TPower and rifle analysis of the PBMC scRNA-seq data following the procedure detailed in Section~\ref{sec:real_design}. The table lists the number of the 1,000 genes in the input data matrix that were assigned positive or negative loadings. Since standard PCA produces non-zero loadings for all genes, the magnitude of the difference between the PCA counts and the counts for the other methods reflects the relative sparsity of the solutions. The "sig. GO" columns capture the number of Gene Ontology Biological Process terms that were significantly enriched among the genes with either positive or negative loadings at an FDR of $\leq 0.1$.
The relative reconstruction error captures average out-of-sample reconstruction error measured over 10 random splits of the data relative to the error for standard PCA (standard deviation is included in parentheses). For each split, PCs were estimated using each method on half of the data and the squared Frobenius norm of the residual matrix for the first two PCs was computed for the other half of the data.}
  \label{table:pbmc_results}
  \end{table}


\begin{table}
\tiny
\centering
 \begin{tabular}{l  | l | r | r | r}
   & Gene Ontology term & \# genes & \# genes  & FDR \\
   & & in term & in group  &  \\   
  \hline
PCA (PC 1, pos) & secretion & 139 & 113 & 1.88e-07 \\
& secretion by cell & 132 & 107 & 1.06e-06 \\  
& vesicle-mediated transport & 172 & 133 & 2.08e-06 \\  
& exocytosis & 101 & 84 & 1.30e-05\\  
& immune effector process & 129 & 102 & 4.95e-05\\  
  \hline  
EESPCA (PC 1, pos) & immune response & 204 & 70 & 8.82e-15\\
& immune system process & 256 & 77 & 4.54e-13 \\
& defense response & 144 & 53 & 5.22e-11 \\
& immune effector process & 129 & 49 & 2.04e-10 \\
& neutrophil activation & 69 & 34 & 6.33e-10\\
  \hline  
EESPCA (PC 2, pos) & adaptive immune response & 51 & 17 & 1.69e-10 \\  
& antigen receptor-mediated signaling pathway & 35 & 12 & 2.11e-06 \\
& antigen processing and presentation of exogenous peptide antigen via MHC...& 16 & 9 & 2.34e-06 \\
& antigen processing and presentation of peptide antigen via MHC class II&17 & 9 & 4.85e-06 \\
& antigen processing and presentation of peptide or polysaccharide antigen...& 17 & 9 & 4.85e-06 \\
  \hline  
EESPCA (PC 2, neg) &regulation of transport & 103 & 23 & 0.0057 \\
& regulation of localization & 154 & 28 & 0.024 \\
& positive regulation of transport & 59 & 16 & 0.027 \\
  \hline   
SPC (PC 1, pos) & secretion & 139 & 111 & 6.94e-08 \\
& vesicle-mediated transport & 172 & 131 & 3.53e-07 \\  
& secretion by cell & 132 & 105 & 4.76e-07 \\  
& exocytosis & 101 & 84 & 7.21e-07\\  
& regulated exocytosis & 97 & 80 & 5.16e-06 \\  
  \hline   
  TPower (PC 1, pos) &  vesicle-mediated transport & 172 & 126 & 8.57e-09 \\  
& secretion & 139 & 105 & 4.50e-08 \\
& secretion by cell & 132 & 100 & 1.23e-07 \\  
& exocytosis & 101 & 80 & 3.79e-07\\  
& immune effector process & 129 & 96 & 2.14e-06 \\  
  \hline  
TPower (PC 2, pos) & adaptive immune response & 51 & 27 & 0.0083 \\
& humoral immune response & 22 & 14 & 0.020 \\
& B cell receptor signaling pathway & 8 & 8 & 0.044 \\  
  \end{tabular}
  \caption{Gene Ontology (GO) enrichment results for positive and negatives loadings on first two PCs generated by standard PCA, EESPC, SPC, and TPower analysis of the PBMC scRNA-seq data. The top five GO Biological Process terms with an FDR of $\leq 0.1$ are listed. Similar results for EESPCA.cv, SPC.1se and rifle are included Table S1 in the Supplementary Results.}
  \label{table:go_results}
  \end{table}

\clearpage

\section{Conclusions}\label{sec:conclusion}

The EESPCA method is a novel sparse PCA technique based on an approximation of the recently rediscovered formula for calculating eigenvectors from eigenvalues for Hermitian matrices \citep{Denton_2021}. In this paper, we have explored the performance of two versions of this method: EESPCA and EESPCA.cv. The EESPCA version uses a fixed threshold of $1/\sqrt{p}$ to induce sparsity in the estimated PC loadings and the EESPCA.cv version selects the threshold via cross-validation to minimize the out-of-sample reconstruction error. 
Compared to the state-of-the-art SPC (and SPC.1se variant) \citep{Witten:2009tg}, TPower \citep{10.5555/2567709.2502610} and rifle \citep{https://doi.org/10.1111/rssb.12291} methods, the EESPCA version can more accurately estimate population PC sparsity across a range of covariance structures, sample sizes and variable dimensionality at a dramatically lower computational cost.
Importantly, EESPCA provides improved estimation of PC sparsity and lower computational cost with an out-of-sample reconstruction error and PC estimation error close to the error achieved by the TPower method and significantly below the error for the SPC, SPC.1se and rifle techniques. In case of limited PC sparsity, i.e, when the proportion of PC loadings with a 0 population value is less than 0.5, the classification performance, out-of-sample reconstruction error and PC estimation error for the EESPCA method deteriorate significantly. In this scenario, the EESPC.cv version provides the best sparsity estimation with PC estimation error and reconstruction error close to the error generated by TPower.
To select the most appropriate sparse PCA method for a given analysis problem, users are encouraged to review the classification performance, PC estimation error, out-of-sample reconstruction error and execution time results for the four different simulation models explored in this paper.
In general, for analysis problems where significant sparsity is assumed for the population PCs, EESPCA should be preferred over EESPCA.cv, SPC, SPC.1se, TPower and rifle. EESPCA is also generally preferable for computationally demanding problems, such as the analysis of very large data matrices or the use of statistical methods like resampling that require repeated application of sparse PCA (if $p \gg n$, the computational cost of SPC may be lower than EESPCA).
If significant sparsity cannot be assumed for the population PCs and computational speed is not critical, then the EESPCA.cv version should be used over EESPCA.
If minimization of PC estimation error or out-of-sample reconstruction error is the key goal and computational speed is not a major concern, then the TPower method provides the best overall performance.

\section*{Acknowledgments}

This work was funded by National Institutes of Health grants K01LM012426, R21CA253408, P20GM130454 and P30CA023108.
We would like to acknowledge the supportive environment at the Geisel School of Medicine at Dartmouth where this research was performed.

%
%
%
%

\bibliographystyle{natbib}
\bibliography{EESPCA_arXiv_v3.bib}
\end{document}